\documentclass[aip,apl,amsmath,amssymb,reprint,]{revtex4-2}
\usepackage{graphicx}
\usepackage{epstopdf}
\usepackage{dcolumn}
\usepackage{bm}
\usepackage{color}
\usepackage{amsmath}

\begin{document}

\title{Infrared full-Stokes polarimetry by parametric up-conversion}

\author{Zhanghang Zhu$^\dag$}
\affiliation{The Key Laboratory of Weak-Light Nonlinear Photonics, Ministry of Education, School of Physics and TEDA Applied Physics Institute, Nankai University, Tianjin, 300071, P.R. China}

\author{Di Zhang$^\dag$}
\affiliation{The Key Laboratory of Weak-Light Nonlinear Photonics, Ministry of Education, School of Physics and TEDA Applied Physics Institute, Nankai University, Tianjin, 300071, P.R. China}

\author{Fei Xie}
\affiliation{The Key Laboratory of Weak-Light Nonlinear Photonics, Ministry of Education, School of Physics and TEDA Applied Physics Institute, Nankai University, Tianjin, 300071, P.R. China}

\author{Jiaxin Chen}
\affiliation{The Key Laboratory of Weak-Light Nonlinear Photonics, Ministry of Education, School of Physics and TEDA Applied Physics Institute, Nankai University, Tianjin, 300071, P.R. China}

\author{Shengchao Gong}
\affiliation{The Key Laboratory of Weak-Light Nonlinear Photonics, Ministry of Education, School of Physics and TEDA Applied Physics Institute, Nankai University, Tianjin, 300071, P.R. China}

\author{Wei Wu}
\affiliation{The Key Laboratory of Weak-Light Nonlinear Photonics, Ministry of Education, School of Physics and TEDA Applied Physics Institute, Nankai University, Tianjin, 300071, P.R. China}

\author{Wei Cai}
\affiliation{The Key Laboratory of Weak-Light Nonlinear Photonics, Ministry of Education, School of Physics and TEDA Applied Physics Institute, Nankai University, Tianjin, 300071, P.R. China}

\author{Xinzheng Zhang}
\affiliation{The Key Laboratory of Weak-Light Nonlinear Photonics, Ministry of Education, School of Physics and TEDA Applied Physics Institute, Nankai University, Tianjin, 300071, P.R. China}

\author{Mengxin Ren}
\email{ren$_$mengxin@nankai.edu.cn}
\affiliation{The Key Laboratory of Weak-Light Nonlinear Photonics, Ministry of Education, School of Physics and TEDA Applied Physics Institute, Nankai University, Tianjin, 300071, P.R. China}
\affiliation{Collaborative Innovation Center of Extreme Optics, Shanxi University, Taiyuan, Shanxi 030006, P.R. China}

\author{Jingjun Xu}
\email{jjxu@nankai.edu.cn}
\affiliation{The Key Laboratory of Weak-Light Nonlinear Photonics, Ministry of Education, School of Physics and TEDA Applied Physics Institute, Nankai University, Tianjin, 300071, P.R. China}

\begin{abstract}

Polarimetry aims to measure polarization information of an optical field, providing a new freedom to enhance performance of optical metrology. Furthermore, the polarimetry in infrared (IR) range holds a promise for a wide range of academic and industrial applications because the IR light relates to unique spectral signatures of various complex chemicals. However, a primary challenge in IR applications remains lacking efficient detectors. Motivated by such a constraint, we present in this paper a nonlinear up-conversion full-Stokes IR Retrieval Polarimetry (IRP) that is able to decipher the IR polarization using a well-commercial and high-performance visible light detector. Assisted by a nonlinear sum-frequency generation (SFG) process in a lithium niobate thin film, the polarization states of the IR light are encoded into the visible SF wave. Based on a Stokes-Mueller formalism developed here, the IR polarization is successfully retrieved from SF light with high precision, and polarization imaging over the targets with either uniform or non-uniform polarization distributions are demonstrated. Our results form a fresh perspective for the design of novel advanced up-conversion polarimeter for a wide range of IR metrological applications. 
\newline

\end{abstract}
\maketitle

Infrared (IR) light is electromagnetic radiation with wavelength longer than that of visible (VIS) light.\cite{Stuart2004infrared} Although invisible to human naked eyes, the IR light holds significant usefulness in various cutting-edge applications including biological sensing,\cite{Rodrigo2015mid} astronomical observation,\cite{Kloppenborg2010infrared} and environmental science,\cite{Pfeilsticker2003atmospheric} owing to its capacity for unique chemical specificity. Until now, the IR related applications are mainly challenged by the lack of efficient detectors.\cite{Keyes2013optical} For example, becuase the IR sensitive semiconductors have small band gap energy, their performance is easily disturbed by ambient thermal fluctuation in terms of dark current and noise. Furthermore, the available pixel number of the arrayed IR detectors is limited. This renders the IR cameras a poor resolution compared with their visible counterparts, such as complementary metal-oxide semiconductors (CMOS) and charge-coupled devices (CCD).\cite{Rogalski2002infrared}

An appealing solution to above constraints is parametric frequency up-conversion, such as sum-frequency generation (SFG).\cite{Barh2019parametric} In such architecture, instead of directly detecting the IR fundamental frequency (FF) photons, the IR signal (at frequency $\omega_1$) is firstly transferred to the visible range with the help of another pump light ($\omega_2$) via a second order nonlinear ($\chi^{(2)}$) material.\cite{Shen1984principles} Then the generated SF photons ($\omega_3=\omega_1+\omega_2$, following the energy conservation law) are detected by high-performance visible detectors. The first study of such parametric up-conversion of IR light could be traced back to 1960s, when Midwinter and Warner demonstrated a conversion of 1.7~$\mu$m radiation to 493 nm by a lithium niobate (LN) crystal.\cite{Midwinter1967up} One year later, such up-conversion technique was further applied to realize IR up-conversion imaging.\cite{Midwinter1968image} Over the past decades, the parametric up-conversion detection has received great interests, and several technical improvements were made in various aspects, such as finding new nonlinear materials for higher up-conversion efficiency,\cite{Zhou2016orbital} ameliorating sensitivity of the visible detectors,\cite{Boyle1970charge} manufacturing high power IR sources at various wavelengths for target illumination,\cite{Hugi2012mid,Schliesser2012mid} optimizing system designs for better imaging field-of-view and resolution.\cite{Torregrosa2015intra,Junaid2019video}

\begin{figure*} [t]
\includegraphics[width=180mm]{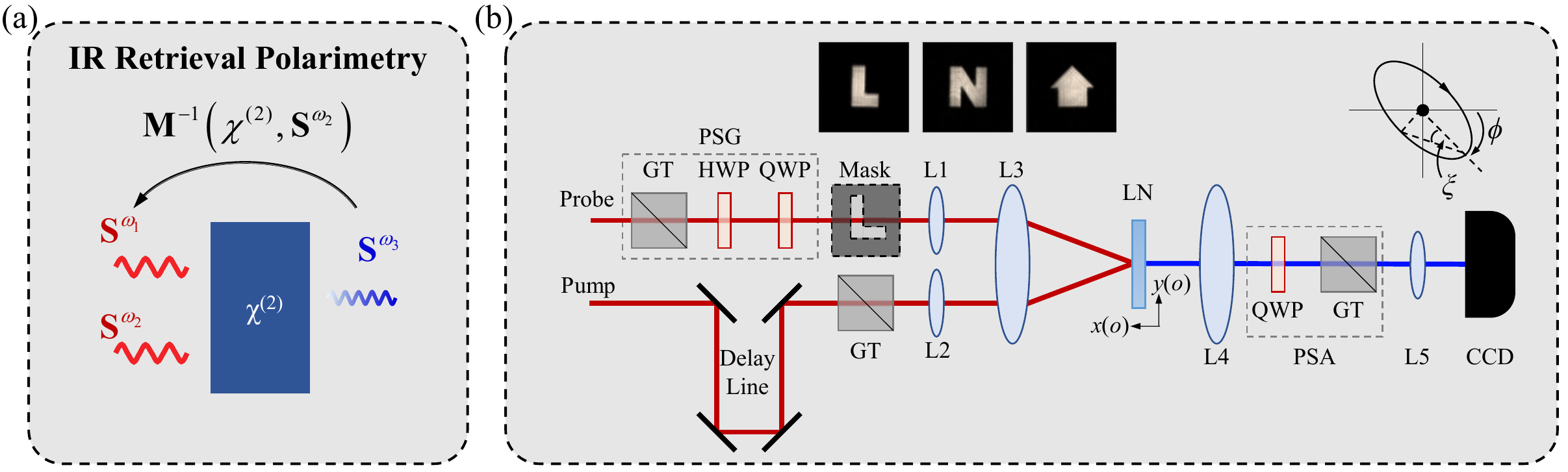} \caption{\label{fig1}
\textbf{A schematic of IRP system.} \textbf{(a)} Operation principle of the IRP system. FF signal ($\omega_1$) and pump light ($\omega_2$) are up-converted into SF wave ($\omega_3$) through a second order nonlinear material. The nonlinear SFG process can be presented by a Mueller matrix $\mathbf{M}\left(\chi^{(2)},\mathbf{S}^{\omega_2}\right)$, via which the Stokes vectors of $\mathbf{S}^{\omega_1}$ can be calculated reversibly from $\mathbf{S}^{\omega_3}$. \textbf{(b)} A sketch of the IRP setup. An optical pump-probe system was built. The path difference between the pump and probe pulses is compensated by a mechanical delay line. The pump polarization is vertically polarized by a Glan-Taylor (GT) prism, and the polarization of the probe light is prepared by a polarization state generator (PSG) consisting of a GT polarizer, a half-wave plate (HWP) and a quarter-wave plate (QWP). The optical beams are focused  onto the LN film respectively afterwards by lens pairs with focal lengths of 500~mm (L1, L2) and 100~mm (L3). The generated SF light is collected by a lens (L4, 75~mm), followed by a polarization state analyzer (PSA) comprising a rotating QWP, and a GT prism, and finally imaged onto a visible CCD camera after a lens (L5, 200~mm). Physical orientation and the shape of the polarization ellipse are defined by angles of azimuth ($\phi$) and ellipticity ($\xi$), respectively. Positive values of $\phi$ and $\xi$ correspond to the clockwise rotation of the polarization azimuth and a right-handed polarization ellipse, as observed against the propagation direction. Intensity images by the signal light for `L', `N' and `house' masks are given in the top-right corner.}
\end{figure*}

However, the current IR up-conversion techniques are still designed to acquire the intensity and spectral information of the IR light, while the polarization information is simply dropped. Complementary to the intensity and spectrum, the polarization is another fundamental characteristic that describes how the light's electric field vector oscillates. While the intensity and spectrum tell us about the geometries and material compositions, polarization interrogation can bring numerous additional richness from the targets. Specifically, by analyzing the light polarization changes by scattering or transmitting through the objects, people can infer birefringence or stress inside material volumes, and roughness or texture of a surface, moreover, enhance contrast for objects that are difficult to distinguish otherwise.\cite{Tyo2006review,Svirko2000Polarization} In this context, polarimetry, as a technique invented to measure the polarization state of light, has become an emerging technique to enhance many fields of optical metrology, optical sensing, and so on.\cite{Rubin2019matrix} 

Enlightened by the respective merits of the two techniques of the nonlinear IR up-conversion and the polarimetry, it would be a good idea to further combine them together. Such incorporation could inherit advantages of both techniques, i.e. detecting IR polarization information while using the VIS detectors. This would consequently benefit many fields of IR related applications. But to the best of our knowledge, such combination still remains unexplored. In this paper, we experimentally demonstrate an IR Retrieval Polarimetry (IRP) that manages to decipher the polarization information of the IR signal from the generated SF light. We develop a Stokes-Mueller formalism based on nonlinear optics that correlates the input IR light polarization states with the SF light. In such algorithmic framework, the incoming IR signals and outgoing SF radiations are represented by 4$\times$1 Stokes vectors, and the nonlinear SFG process is represented by a 4$\times$4 Mueller matrix. We further experimentally achieve the IR polarization imaging of different targets with uniform or non-uniform polarization distributions using our IRP system, which reconstructs the polarization of the IR signal light with high precision. Such IRP technique would pave the way to the design of novel IR polarimetry for advanced remote sensing, astronomy and industrial inspection.

\section{Stokes-Mueller formalism for IRP}
First, we develop a Stokes-Mueller algebra for retrieving the polarization of the IR signal light based on that of the SF wave, which essentially forms the mathematical basis for the operation of our IRP. We adopted here a four-element Stokes vector $\mathbf{S}=[s_0, s_1, s_2, s_3]^T$ ($^T$ denotes matrix transpose) to describe the polarization state of light.\cite{Schmieder1969stokes} The Stokes description of light has a chief advantage that it operates in the form of intensity, which is real valued and is easily measurable using photo-detectors in experiments. Furthermore, in contrast to other representations, such as the Jones vector that is limited to the case of fully polarized light,\cite{Bass2010} the Stokes vector also spans the space of unpolarized and partially polarized light. Each Stokes element is defined as a sum or difference between intensities of different orthogonal polarization bases. Specifically, $s_0$ is the total intensity of the light, $s_1$ is the intensity difference between linear polarization components along 0$^\circ$ and 90$^\circ$, $s_2$ is the intensity difference between $+45^\circ$ and $-45^\circ$ linear polarizations, and $s_3$ is the intensity difference between right- and left-handed circular polarization components. In such representation, the polarization dependent interaction between the light and matter is denoted by a 4$\times$4 Mueller matrix ($\mathbf{M}$), and the incident polarization state $\mathbf{S}$ relates to the output state $\mathbf{S'}$ via $\mathbf{S'}=\mathbf{M}\cdot\mathbf{S}$. Such Stokes-Mueller formalism acts as a basis for the traditional linear optical polarimetry to characterize the signal light polarization.\cite{Bass2010} Here we further develop a new nonlinear Stokes-Mueller formalism for the IRP technique, which aims to reconstruct the IR polarization from that of the SF light, rather than measuring that of the IR light directly, as illustrated by Fig.~1(a).

\begin{figure*} [htp]
\includegraphics[width=141mm]{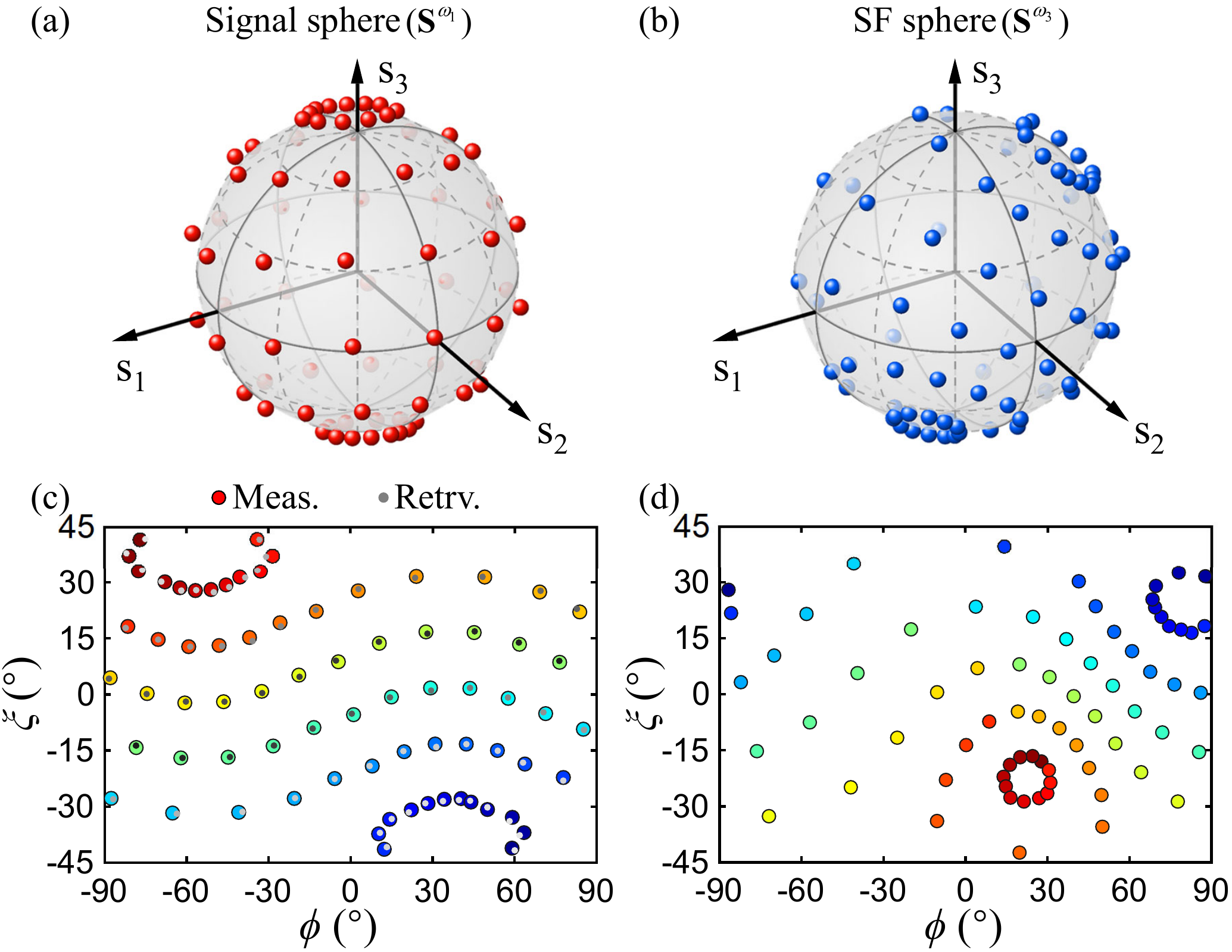} \caption{\label{fig2}
\textbf{Calibration of IRP.} \textbf{(a)} Poincar\'e sphere of FF signal light (at $\omega_1$). 72 different signal polarization states (red dots) that uniformly sample the sphere were chosen as input polarizations for calibrating IRP. \textbf{(b)} Poincar\'e sphere of generated SF light. \textbf{(c)} The three-dimensional FF Poincar\'e sphere is expanded into a planar map in a coordinate of azimuth--ellipticity angles ($\phi-\xi$). The originally input FF polarization states are presented by the dots in color, and the reconstructed results are given by small gray dots. The well overlap between them demonstrates the high precision of our IRP in reconstructing the FF polarizations. \textbf{(d)} The $\phi-\xi$ map of the SF Poincar\'e sphere. The SF dots generated by the FF light are shown using the same color as those in (c).}
\end{figure*}

According to the nonlinear optics, the generated SF waves are closely related to the second order nonlinear polarizations $\mathbf{P}^{(2)}$ excited inside the materials. Specifically, the $\mathbf{P}^{(2)}$ for the SFG process follows\cite{Boyd2003nonlinear}
\begin{equation}
\begin{small}
	P^{(2)}_{i}(\omega_3)=2\sum_{j,k}\epsilon_0\chi_{ijk}^{(2)}E_j^{\omega_1}E_k^{\omega_2},
\end{small}
\end{equation}
\noindent in which footnotes $i,j,k$ stand for the unit vector along the Cartesian $x$-, $y$- or $z$-axes, and $E_j^{\omega_1}$ and $E_k^{\omega_2}$ are the electric fields of the signal and pump light, respectively. $\chi_{ijk}^{(2)}$ is the element of the nonlinear susceptibility tensor, which generates $\mathbf{P}^{(2)}$ vector polarized along the $i$-direction under the combined interaction of the fundamental electric fields along $j$- and $k$-directions. Such $P^{(2)}_{i}$ acts as a secondary source to radiate the nonlinear SF waves $\mathbf{E}^{\omega_3}$ polarized along the $i$-direction, and magnitude of the $\mathbf{E}^{\omega_3}$ is proportional to $\mathbf{P}^{(2)}$. By further considering the mathematical relationship between $\mathbf{S}^{\omega_3}$ and $\mathbf{E}^{\omega_3}$, we could derive

\begin{equation}
\mathbf{S}^{\omega_3}=\mathbf{M}\left(\chi^{(2)}, \mathbf{S}^{\omega_2}\right)\cdot\mathbf{S}^{\omega_1},
\end{equation}

\noindent in which $\mathbf{M}\left(\chi^{(2)},\mathbf{S}^{\omega_2}\right)$ is a 4$\times$4 Mueller matrix describing the SFG process. Equation~(2) has a similar form as the linear Stokes-Mueller formalism. However, different from the Mueller matrix in linear optics that only relates to medium's properties such as birefringence or optical chirality,\cite{Bass2010} the SFG matrix $\mathbf{M}$ here is a function of not only properties of the material (i.e. nonlinear $\chi^{(2)}$), but also the pump polarization $\mathbf{S}^{\omega_2}$. Once the SFG matrix $\mathbf{M}$ and Stokes parameters of the SF light ($\mathbf{S}^{\omega_3}$) are determined, the polarization state of the signal light could be calculated via

\begin{equation}
\mathbf{S}^{\omega_1}=\mathbf{M}^{-1}\left(\chi^{(2)}, \mathbf{S}^{\omega_2}\right)\cdot\mathbf{S}^{\omega_3}.
\end{equation}

\section{Setup of IRP}
An $x$-cut LN thin film was used as the SFG material in our experiment. The thickness of the LN film is about 200~nm, which resides on a 500~$\mu$m thick silicon dioxide (SiO$_2$) substrate (NANOLN, Jinan Jingzheng Electronics Co., Ltd). LN is a typical optical material showing ultra-broadband transparency and strong second order nonlinearity covering the visible to mid-IR spectral range, which renders it a feasible material platform for wide photonic and optoelectronic applications. The LN crystal belongs to a point group of $3m$. Resulting from the symmetry restrictions, $\chi^{(2)}$ tensor of the LN has 11 non-vanishing components, in which only 4 elements are independent: $\chi^{(2)}_{eee}$, $\chi^{(2)}_{eoo}$, $\chi^{(2)}_{ooo}$, $\chi^{(2)}_{ooe}$.\cite{Ma2020second} The experimental axes were defined to overlap with the LN principal crystallographic coordinate. Optical axis of the LN is nominated as $z(e)$-axis, and the light propagated along $-x$-direction, as shown in Fig.~1(b). 

\begin{figure*} [htp]
\includegraphics[width=150mm]{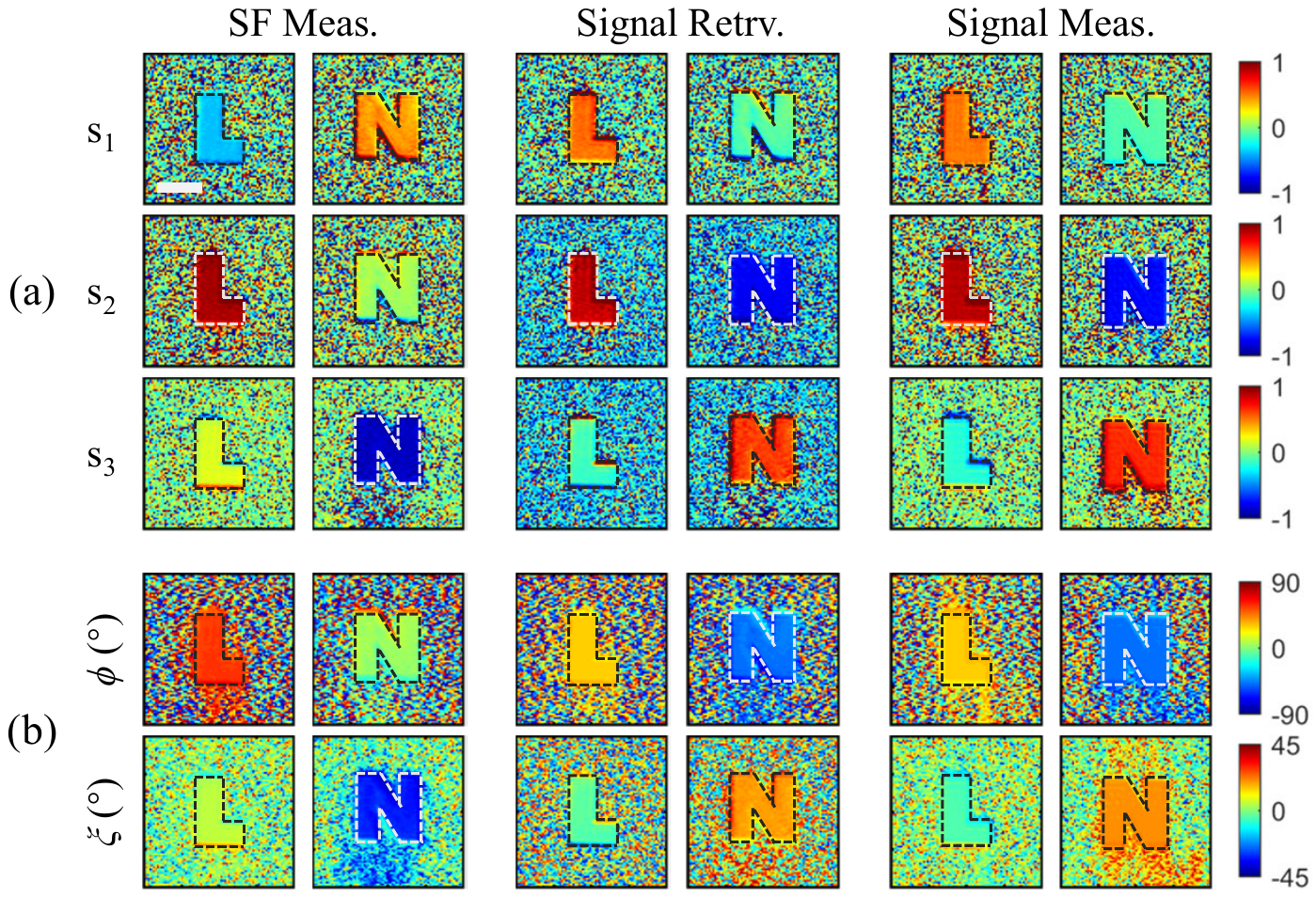}\caption{\label{fig3}
\textbf{Polarization images of `L' and `N'.} \textbf{(a)} Images of Stokes parameters $s_1$, $s_2$, and $s_3$. The first, second and third columns give the experimentally measured results of SF waves, the retrieved FF signal, and the directly measured FF results, respectively. Outlines of the letters are sketched by dashed lines. Scale bar of 100~$\mu$m is given in bottom-left conner of the first picture. \textbf{(b)} Images are given in forms of azimuth ($\phi$) and ellipticity ($\xi$) of the polarization ellipse. It is clear that the retrieved FF polarization parameters reproduce the directly measured FF results very well, confirming the precision of our technique.}
\end{figure*}

A schematic of the experimental setup is shown in Fig.~1(b). An optical pump-probe system was built. The near-infrared (NIR) laser pulses at a wavelength of 808~nm were generated by a Kerr lens mode-locked Ti:sapphire laser oscillator (Spectra Physics, Maitai, pulse width of 230~fs, repetition rate of 80~MHz). The pulse train was divided into two paths using a beam splitter. A delay line, consisting of a linear translation stage in the pump beam path, was used to compensate the length difference between the pump and probe paths and ensured the fine control of the pulse synchronization. The pulses were adjusted to be well overlapped temporally to guarantee the strongest SFG. The optical beams were focused onto the LN film afterwards by lens pairs with focal lengths of 500~mm and 100~mm, forming focal points with the diameter of 135~$\mu$m. The intensity of the pump and probe beams were about 142~MW/cm$^2$ and 130~MW/cm$^2$ on the LN surface, respectively. The polarization of the pump beam was fixed along the ordinary ($o$-) axis of the LN crystal. The generated SF light at 404~nm was collected in the forward direction by a lens with the focal length of 75~mm. Short-pass filters (BG40, not shown) were inserted in the SF emission path to block the remaining 808~nm light. The SF light then entered a home-built full-Stokes polarization state analyzer (PSA) comprising a rotating quarter-wave plate (QWP), a Glan-Taylor (GT) calcite polarizer, and afterwards was detected by a monochrome electron-multiplying charge-coupled device (EMCCD, iXon Ultra 888). 91 frames of SF signal were recorded while the QWP was rotated from 0 to 180$^\circ$ with a discrete step of 2$^\circ$. Through Fourier transformation of the recorded SF intensity as a function of the QWP rotation angle, the Stokes parameters of the SF wave could be obtained.\cite{Ren2015linearly} 

\section{Calibration of IRP}
In order to accurately reconstruct $\mathbf{S}^{\omega_1}$ of the incident IR signal light via the measured SF Stokes parameters according to Eq.~(3), the first objective is to determine the Mueller matrix $\mathbf{M}\left(\chi^{(2)},\mathbf{S}^{\omega_2}\right)$ of the IRP system. Such process could be referred as calibration of the IRP, which could be accomplished by resorting to multiple sets of known input and output polarization pairs. 

The polarization states of the signal beam incident on the LN were prepared by a polarization state generator (PSG) consisting of a GT polarizer, a half-wave plate (HWP) and a QWP. For the calibration end, 72 different signal polarization states that uniformly sample the Poincar\'e sphere were chosen as input signal polarizations, whose $\mathbf{S}^{\omega_1}$ parameters were further experimentally characterized using the PSA [as illustrated by red dots in Fig.~2(a)]. The corresponding generated $\mathbf{S}^{\omega_3}$ were measured as well and labeled by blue dots on the SF Poincar\'e sphere in Fig.~2(b). The 16 elements of $\mathbf{M}$ linking $\mathbf{S}^{\omega_1}$ to $\mathbf{S}^{\omega_3}$ can be numerically determined using a least squares method. Afterwards, the FF polarization states can be retrieved through Eq.~(3). For the clearer comparison between the original and the retrieved IR signal polarizations, the three-dimensional Poincar\'e spheres are expanded into planar maps, as shown in Fig.~2(c,d), in forms of angles of azimuth [$\phi=\frac{1}{2}\arctan(\frac{s_2}{s_1}$)] and ellipticity [$\xi=\frac{1}{2}\arctan(\frac{s_2}{\sqrt{s_1^2+s_2^2}}$)], which correspond to the physical orientation and the shape of the polarization ellipse, respectively. The well overlap between the reconstructed polarization states (small gray dots) and the directly measured ones (large color dots) in Fig.~2(c) demonstrates the excellent precision of our IRP in reconstructing FF signal polarization states from the SF polarizations. 

\begin{figure} [htp]
\includegraphics[width=75mm]{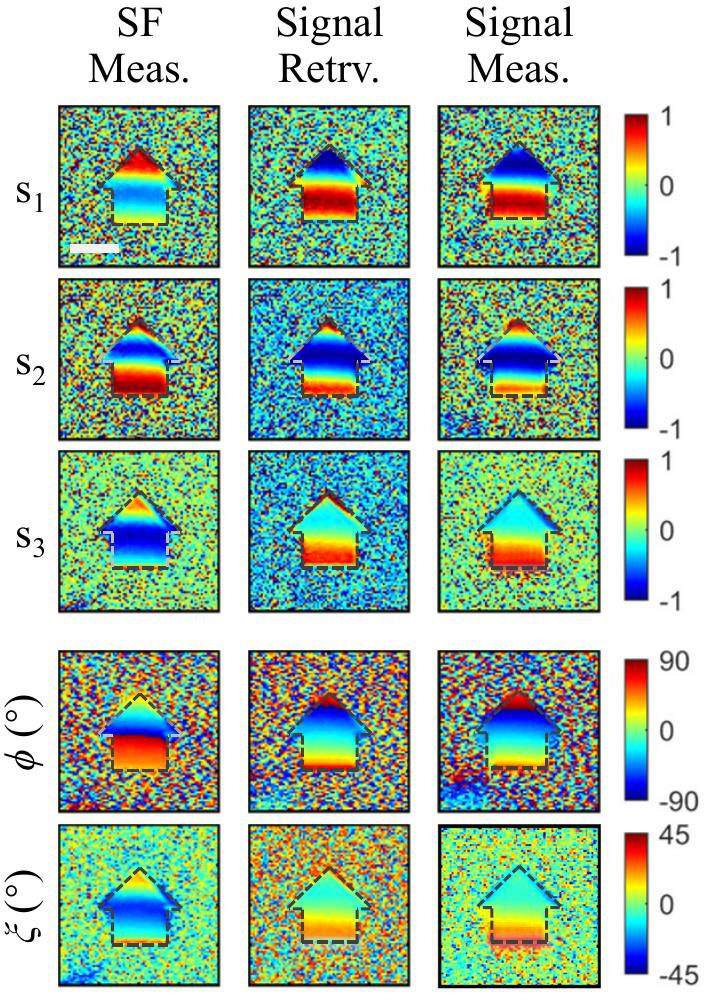} \caption{\label{fig4}
\textbf{Polarization images of a `house' mask with a non-uniform birefringence distribution.} The images are given in forms of Stokes parameters ($s$), azimuth ($\phi$) and ellipticity ($\xi$). The retrieved FF signal polarization images (the second column) reproduce the measured FF values (the third column) very well.}
\end{figure}

\section{Stokes polarimetric imaging}
We now apply our IRP in full-Stokes polarization imaging. Such polarization imaging technique aims to map the polarization states across a scene of interest. In the traditional IR imaginary polarimetry, the polarization distribution over the cross-section of the signal beam is directly characterized by the PSA equipped by a IR camera. However, here we would show the retrieval of the spatial polarization information contained in the IR signal beam via analyzing that of the SF wave. To demonstrate such application, binary masks with transparent letters of `L' and `N' were used as the targets, which were fabricated by focused ion beam milling through a 200~nm thick opaque gold film supported by a SiO$_2$ substrate [the intensity images of the masks are given in Fig.~1(b)]. The masks were inserted into the signal beam, and were projected onto the LN film plane. The polarization states of the signal light illuminating the mask were adjusted by the PSG in front. In this configuration, the polarization states across the signal beam were uniformly distributed. After interacting with the pump beam, both the morphological and polarization information of the masks contained in the signal beam were successfully transferred into the generated SF light and recorded by the PSA. The SF Stokes parameters were analyzed at each pixel, and the polarization images at the SF wavelength were constructed afterwards. Figure 3 presents the results of the masks by our IRP system. In the literatures of polarization imaging, different sets of polarization parameters are commonly used. The first one is the Stokes parameters ($s$). In contrast to the traditional intensity photography that gives the same images despite the signal light polarization varies, the $s_1$, $s_2$ and $s_3$ present a good capability in distinguishing different signal polarizations. The Stokes parameters are uniformly distributed inside the letter area, while the background is noisy because no reliable SF signal was detected resulting from the opacity outside the letter areas. For each mask, the Stokes images at the SF wavelength are given in the first column. The retrieved FF images following Eq.~(3) are given in the second column, and the directly measured FF results are given in the third column. It is clear that the retrieved FF polarizations reproduce the directly measured FF values very well, indicating the fidelity and precision of our technique in reconstructing the FF polarization images. Images of $\phi$ and $\xi$ are further plotted in Fig.~3(b) to intuitively present the geometric features of the polarization ellipse, which also confirm the good consistence between the retrieved and the directly measured FF signal values.

We further implement our IRP technique to the target that contains an inhomogeneous birefringent structure, which causes multiple polarization components in the signal light. Such scenario holds a high application significance in the fields of remote sensing for mineral exploration, as well as microscopy for characterizing the anisotropic biological tissues. To demonstrate such functionality, a commercial depolarizer was adopted as the target. A `house' shaped mask was adopted [as shown by the intensity image in Fig.~1(b)] and put on the surface of the depolarizer. Such depolarizer is a patterned microretarder array, which consists of a thin film of birefringent liquid crystal polymer sandwiched between two glass plates. The fast-axis orientation and phase retardation of the liquid crystal were designed to vary with a periodical pattern inside the depolarizer. Thus, after transmitting through such depolarizer, a spatially varied polarization distribution would be produced over the cross section of the signal light.  It is clear that the birefringence feature of the depolarizer is not discernible by the intensity image in Fig.~1(b), however, it is very well resolved in the polarization images in Fig.~4. The pseudo-color plot essentially traces the contour of the spatial birefringence variation of the liquid crystal, and each band of color in the polarization images represents a region with the same polarization state. Similar as Fig.~3, the retrieved FF polarization images here (the second column of Fig.~4) reproduce the directly measured FF values (the third column) very well, confirming the precision of our technique again. 

\section{Conclusion}
In conclusion, we build a nonlinear up-conversion full-Stokes IRP that manages to retrieve the polarization information of the incident IR probe light from that of the up-converted SF wave. The IR polarization images of targets with either uniform or non-uniform polarization distributions are successfully reconstructed using our IRP system with high precision. Despite we focus here on the degenerate SFG process that both the pump and probe light have the same wavelength at NIR, it is expected that our method also validates for the non-degenerate SFG process where the signal photon can be in mid- or long-infrared range and up-converted by visible pump light.\cite{Falk1971detection,Kviatkovsky2020microscopy} Thus it is applicable to vibrational spectroscopy for birefringent molecule identification, and thermal imaging for night vision detection. Furthermore, utilizing the current state-of-the-art scheme of nonlinear enhancement by nanostructures such as metasurfaces\cite{Li2017nonlinear,Ma2021nonlinear,Fedotova2020second,Carletti2021steering} or intracavity up-conversion configuration,\cite{Dam2012room,Israelsen2019real} the efficiency of the SFG process from the LN film could be furthermore effectively improved. We could foresee a direct impact of our results on a variety of IR polarimetric applications, such as optical crystallography, chemical sensing, disease diagnosis, and explosive detection, etc.

\begin{acknowledgments}
Z.Z. and D.Z. contribute equally to this work. This work was supported by National Key R\&D Program of China (2017YFA0305100, 2017YFA0303800, 2019YFA0705000); National Natural Science Foundation of China (92050114, 91750204, 61775106, 11904182, 12074200, 11774185); Guangdong Major Project of Basic and Applied Basic Research (2020B0301030009); 111 Project (B07013); PCSIRT (IRT0149); Open Research Program of Key Laboratory of 3D Micro/Nano Fabrication and Characterization of Zhejiang Province; Fundamental Research Funds for the Central Universities (010-63201003, 010-63201008, 010-63201009, 010-63211001); Tianjin Youth Talent Support Program. We thank Nanofabrication Platform of Nankai University for fabricating samples.
\end{acknowledgments}

\section*{References}
\bibliography{Article_file_v23.bbl}

\end{document}